\documentclass[letterpaper,12pt,copyedit]{article} %% double spaced
\usepackage{osajnl2} %% do not use with REVTeX4
\usepackage[draft]{hyperref} %% optional
\usepackage{amsmath, amssymb, amsthm}
\begin{document}

\title{Photonic Bloch-dipole-Zener Oscillations in Binary Parabolic Optical Waveguide Arrays}

\author{Ming Jie Zheng,$^{1,*}$ Yun San Chan,$^1$ and Kin Wah Yu$^{1,2}$}
\address{$^1$Department of Physics, The Chinese University of Hong Kong, Shatin, New Territories, Hong Kong, China}
\address{$^2$Institute of Theoretical Physics, The Chinese University of Hong Kong, Shatin, New Territories, Hong Kong, China}
\address{$^*$Corresponding author: mjzheng@phy.cuhk.edu.hk, current email: mzheng@cae.wisc.edu.}

\begin{abstract}

We have studied the propagation and Zener tunneling of light in the binary parabolic optical waveguide array (BPOWA), which consists of two evanescently coupled dissimilar optical waveguides. Due to Bragg reflections, BPOWA attains two minibands separated by a minigap at the zone boundary. Various coherent superpositions of optical oscillations and Zener tunneling occur for different parameters on the phase diagram. In particular, Bloch-Zener oscillation and a different type of Bloch-dipole-Zener oscillation are obtained by the field-evolution analysis. The results may have potential applications in optical splitting and waveguiding devices and shed light on the coherent phenomena in optical lattices.

\end{abstract}


\maketitle

\newpage

\section{INTRODUCTION}

Bloch oscillation (BO) \cite{ZPhys.52.555.1928}, dipole oscillation (DO) \cite{NewJPhys.5.112.2003}, and Zener tunneling (ZT) \cite{RSocLondA.145.523.1934} are three fundamental and important transport phenomena of particles or waves in periodic potentials of different natures (electronic, plasmonic, photonic, acoustic, matter wave, etc.). BO was first proposed to describe the oscillatory motion of electrons in periodic potential under a constant force  \cite{ZPhys.52.555.1928}. BO cannot persist forever, but decay through dissipation\cite{OptLett.29.2485.2004}, defects \cite{PhysRevB.81.195118.2010}, tunneling \cite{PhysRevLett.96.053903.2006}, and other damping effects. Among them, ZT plays an important role as an interband transition \cite{RSocLondA.145.523.1934}. If the potential is not linear but parabolic, the oscillation at the bottom of the parabolic band is called DO \cite{NewJPhys.5.112.2003, LaserPhys.16.367.2006}. Electronic BO is hard to observe in regular lattices because the coherence time of electrons is shorter than the BO period, and the electro-optic method was proposed to observe long dephasing time of electronic BO \cite{PhysRevB.50.8106.1994}. Photonic BO is more easily observed due to the longer coherence length of optical wavepackets \cite{PhysRevLett91.263902.2003}. In addition to photonic BO \cite{OptLett.23.1701.1998, PhysRevLett.83.4752.1999, PhysRevLett.83.4756.1999, PhysRevLett91.263902.2003}, there are also various analogies of electronic BO, such as plasmonic BO \cite{ApplPhysLett.91.143121.2007, ApplPhysLett.91.243113.2007, ApplPhysLett.94.161105.2009}, acoustic BO \cite{PhysRevLett98.134301}, matter wave BO \cite{NewJPhys.9.62.2007}, etc.

The superposition of BO and ZT, which is referred as Bloch-Zener oscillation (BZO), has been studied theoretically \cite{NewJPhys.8.110.2006, EurophysLett.76.416.2006, PhysRevLett.101.193902.2008} and experimentally \cite{PhysRevLett.96.023901.2006, PhysRevLett.102.076802.2009}. BZO have potential applications in optical beam splitters and interferometers \cite{NewJPhys.8.110.2006, PhysRevLett.102.076802.2009}, and also be applied to obtain different propagation patterns in two layers of optical waveguide ladders \cite{OWL}. Steering between BO and DO has been realized in a parabolic optical waveguide array \cite{JOptSocAmB.27.1299.2010}. To our knowledge, there has been no study on the superposition of DO and ZT up to now. As we know, DO occurs in the parabolic band, and ZT takes place in the multiband system. To connect DO and ZT, we propose to study a binary parabolic optical waveguide array (BPOWA), because it has two parabolic minibands.

In this work, the BPOWA has two minibands with a band gap at the edge of the Brillouin zone. The gap can be tuned to be small enough to make sure the occurrence of ZT, when the difference between the onsite propagation constants of two kinds of waveguides are proper. Various superpositions of optical oscillations and ZT are shown on the phase diagram. The superpositions of BO (approximate), DO and ZT can be realized, which is demonstrated by the field-evolution analysis. When the gap increases, the tunneling rate decreases. The band structure can be demonstrated visually by the spatial evolution of Bloch oscillation. The combination of BO, DO and ZT may have potential applications in optical splitting and waveguiding devices.

\section{MODEL AND FORMULA}

\begin{figure}[htbp]
  \centering
  \includegraphics[width=0.8 \textwidth]{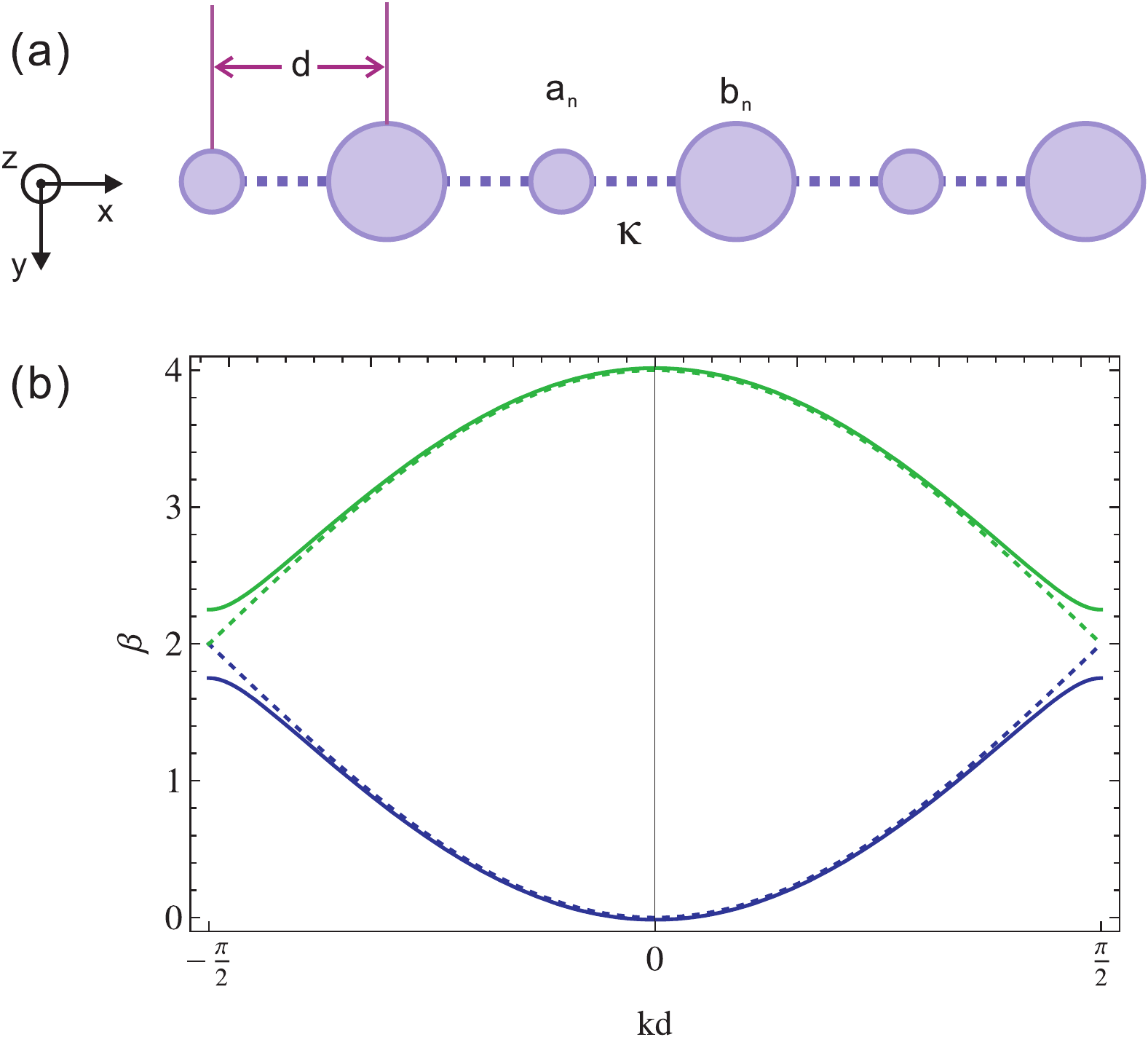}
  \caption{(Color online) (a) Schematic diagram of a binary parabolic optical waveguide arrays (BPOWA). (b) Dispersion relation of unitary POWA (dotted lines) and binary POWA (solid lines). The parameters are $N=101$, $\alpha = 2.0$, $\kappa = 1.0$, and $\delta V = 0.5$. Fig1.eps.}
  \label{fig:BPOWA}
  \end{figure}

\begin{figure}[htbp]
  \centering
  \includegraphics[width=0.8 \textwidth]{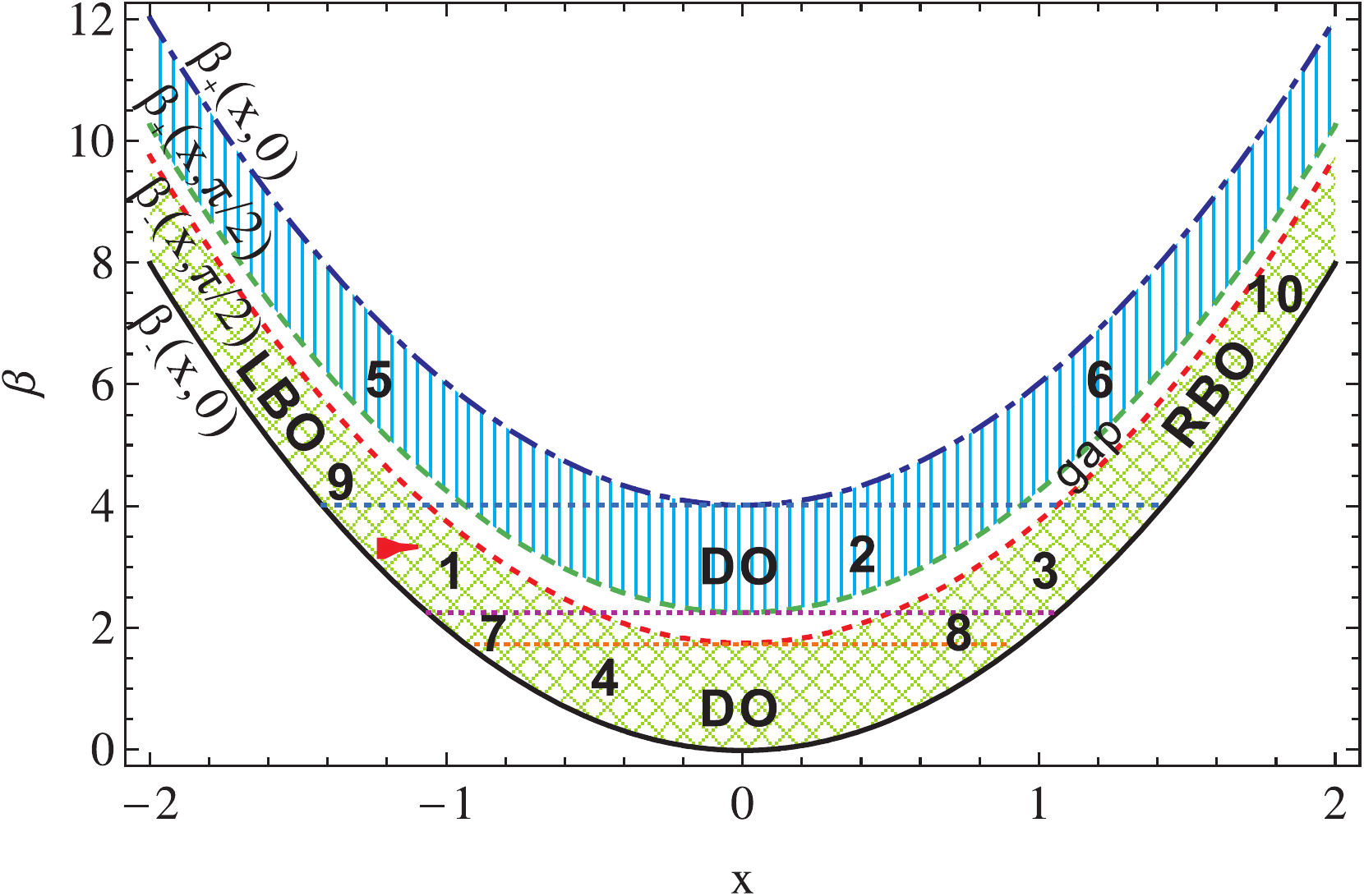}
  \caption{(Color online) Phase diagram of BPOWA. There are two bands, whose boundaries are $\beta_{-}(x, 0)$, $\beta_{-}(x, \pi/2)$, $\beta_{+}(x, \pi/2)$, and $\beta_{+}(x, 0)$. Separated by three critical lines $\beta_{-}(0, \pi/2)$, $\beta_{+}(0, \pi/2)$, and $\beta_{+}(0,0)$, ten regions are formed. The correspondences between these regions and different optical oscillations are as follows. DO (dipole oscillation): 2 and 4. LBO (left Bloch oscillation): 1, 5, 7, and 9. RBO (right Bloch oscillation): 3, 6, 8, and 10. The parameters are $N=101$, $n_0=20$, $k_0 = \pi$, $\sigma=1$, $\alpha = 2.0$, $\kappa = 1.0$, and $\delta V = 0.5$. Fig2.eps.}
  \label{fig:PD}
  \end{figure}

The binary parabolic optical waveguide arrays (BPOWAs) are formed by two types of waveguides placed alternatively with spatial period $2d$, as shown in Fig.~\ref{fig:BPOWA}. We consider the nearest-neighbor coupling with coupling constant $\kappa$. The paraxial propagation of light beam in the BPOWA is described by the evolution equation of the modal amplitudes $a_n$ and $b_n$ as
\begin{equation}\label{eq:Modal}
\begin{aligned}
& \left[\mathbf{i}\frac{d}{dz}+V(x_n) - \frac{\delta V}{2} \right]a_n(z) + \kappa [b_{n-1}(z) + b_{n}(z)] = 0\,, \\
& \left[\mathbf{i}\frac{d}{dz}+V(x_n) + \frac{\delta V}{2} \right]b_n(z) + \kappa [a_{n}(z) + a_{n+1}(z)] = 0\,,
\end{aligned}
\end{equation}
where the parabolic potential profile is $V(x_n) = \alpha x_n^2 + \alpha$, $\delta V$ is the onsite propagation constants difference of two different kinds of waveguides, and $\kappa$ is the nearest-neighbor coupling constant. By substituting the solutions $a_n^m(z)=u_n^m e^{\mathbf{i}\beta_m z}$ and $b_n^m(z)=v_n^m e^{\mathbf{i}\beta_m z}$ into Eq.~(\ref{eq:Modal}), we obtain the eigenvalue equation in matrix form as
\begin{equation}\label{eq:Eigen}
\textbf{H} \left(\begin{array}{c}
              \textbf{u} \\
              \textbf{v}
            \end{array}
      \right)
= \beta \left( \begin{array}{c}
              \textbf{u} \\
              \textbf{v}
            \end{array}
            \right)\,,
\end{equation}
where $\textbf{H} =\{\{ V(x_n) - \frac{\delta V}{2}, \kappa(e^{-\mathbf{i}2kd} +1)\},\{\kappa(e^{\mathbf{i}2kd} +1), V(x_n) + \frac{\delta V}{2}\}\}$. From Eq.~(\ref{eq:Eigen}), the dispersion relation is derived as
\begin{equation}\label{eq:Disp}
\beta_{\pm}(x,k) = V(x) \pm \sqrt{2\kappa^2(1+\cos 2kd) + \frac{\delta V^2}{4} }\,.
\end{equation}
There are two bands in the dispersion relation, as shown in Fig.~\ref{fig:BPOWA}(b) with $x=0$. When $\delta V =0$, there is no gap between the two bands (as shown by the dotted lines). While for finite difference of onsite propagation constants, e.g., $\delta V =0.5$, a gap opens at the edge of the Brillouin zone.
The phase diagram of BPOWA in the $\beta$-$x$ domain is shown in Fig.~\ref{fig:PD}. The two minibands $\beta_{-}$ and $\beta_{+}$ are formed with four boundary lines $\beta_{-}(x, 0)$, $\beta_{-}(x, \pi/2)$, $\beta_{+}(x, \pi/2)$, and $\beta_{+}(x, 0)$. In each band, there exists a critical line $\beta_{-}(0, \pi/2)$ (or $\beta_{+}(0, 0)$), which separates DO and BO. These two critical lines and an additional critical line $\beta_{+}(0, \pi/2)$ separate the two bands into ten different regions marked from 1 to 10, which correspond to various optical oscillations. The correspondences between optical oscillations and phase regions are as follows: DO (dipole oscillation): region 2 and 4. LBO (left Bloch oscillation): region 1, 5, 7, and 9. RBO (right Bloch oscillation): region 3, 6, 8, and 10. The band gap is so small that ZT between two minibands is possible. Five different combinations of optical oscillations are formed: LBO-ZT-LBO (5-9), RBO-ZT-RBO (6-10), LBO-ZT-DO-ZT-RBO (1-2-3), LBO-RBO (7-8), and DO (4). For each combination, the oscillation of light beam can transit from one kind to another through Zener tunneling. In the region 1-2-3, the superposition of BO, DO and ZT is realized, we call such superposition as Bloch-dipole-Zener oscillation (BDZO). For example, a typical input light beam with components as shown by the red profile in region 1 can undergo the BDZO. Due to the involvement of DO, the oscillation spatial range of BDZO is much wider than the usual Bloch-Zener oscillation (BZO). This property of BDZO may have potential application in optical waveguiding devices.

\section{Field evolution of BZO and BDZO}

The propagation of light beam for BZO and BDZO can be demonstrated by the field-evolution analysis. We submit an input Gaussian beam \cite{PhysRevA.81.033829}
\begin{equation}
\psi(n, 0)=\frac{1}{(2\pi\sigma^2)^{1/4}} e^{-\frac{(n - n_0)^2}{4 \sigma^2}}e^{-\mathbf{i}k_0(n - n_0)}\,,
\end{equation}
with transverse wavenumber $k_0$ and intensity profile $|\psi(n,0)|^2$, which has a discrete Gaussian distribution centered at the $n_0$th waveguide with a spatial width $\sigma$. The exponential factor $\exp{[-\mathbf{i}k_0(n - n_0)]}$ captures the phase differences between the input beams excited at the $n$th and the $n_0$th waveguides. The zero phase difference $k_0 = 0$ indicates that the input beam has a plane wavefront.

The input Gaussian beam can be expanded in terms of normal modes as
\begin{equation}\label{eq:expansion}
|\psi (n, 0)\rangle = \sum_{m} A_m |m\rangle\,,
\end{equation}
where $A_m = \langle m|\psi(n,0)\rangle$ is the constituent
component of the input Gaussian beam. The subsequent
wave function at propagation distance $z$ is
\begin{equation}\label{eq:beam}
|\psi (n,z)\rangle = \sum_{m} A_m e^{\mathbf{i}\beta_m z}|m\rangle\,.
\end{equation}
The corresponding wave function in the reciprocal space can be obtained by taking the following Fourier transform
\begin{equation}\label{eq:phik}
|\phi (k,z)\rangle = \mathcal {F}[|\psi (x,z)\rangle]\,.
\end{equation}
For BZO, the components of input beam and field evolution in the spatial and reciprocal space are shown in Fig.~\ref{fig:bxk0} for three different cases: (a)-(c) $\delta V = 0.0$, (d)-(f) $\delta V = 0.5$, and (g)-(i) $\delta V = 1.0$. In each case, the same input Gaussian beam is used, whose components lies in the region 5, which corresponds to LBO. When $\delta V = 0$, that is, the onsite propagation constants of waveguides are the same, the two bands are merged into one. Thus the light beam undergoes LBO between the two boundaries of the merged band. The contour plots of $|\psi(x,z)|^2$ in the spatial space and $|\phi(k,z)|^2$ in the reciprocal space are shown in Figs.~\ref{fig:bxk0}(b) and \ref{fig:bxk0}(c), respectively. The light beam oscillates in the left side of the system in the spatial space and is almost constantly accelerated in the reciprocal space, which are obvious features of LBO. When $\delta V = 0.5$, the band splits into two minibands with a small gap, which opens at the edge of Brillouin zone $kd = \pi/2$. The gap is so small that ZT is possible. As shown in Fig.~\ref{fig:bxk0}(e), the input light beam undergoes LBO at the beginning. Around $kd = \pi/2$, the beam splits into two paths due to ZT. The intensities of two LBO paths are almost the same, and in the reciprocal space, the slanted path of momentum also splits into two when ZT occurs, as shown in Fig.~\ref{fig:bxk0}(f). When $\delta V = 1.0$, the band gap becomes wider, the tunneling rate becomes smaller, that is, the intensity of LBO path in the upper band is stronger than that in the lower band, as shown in Fig.~\ref{fig:bxk0}(h). In the reciprocal space, most of the intensity of the slanted path of momentum is confined in the range from $-\pi/2$ to $\pi/2$, as shown in Fig.~\ref{fig:bxk0}(i).

As mentioned in the description of Fig.~\ref{fig:PD}, if the components of input light beam lies in the region 1 and the band gap is properly wide, it will undergo BDZO in the BPOWA. Figure \ref{fig:bxkPi} shows the components of input light beam and the field evolution in the spatial and reciprocal spaces for three different cases: (a)-(c) $\delta V = 0.0$, (d)-(f) $\delta V = 0.5$, and (g)-(i) $\delta V = 1.0$. When $\delta V = 0$, the two minibands emerge into one, and the light beam behaves like in a unitary POWA. It undergoes DO, that is, oscillatory motion in both the spatial and reciprocal spaces, as shown in Figs.~\ref{fig:bxkPi}(b) and \ref{fig:bxkPi}(c). When $\delta V = 0.5$, the band splits into two minibands with a small gap, as shown in Fig.~\ref{fig:bxkPi}(d). The gap is so small that ZT is possible. The light beam first undergoes LBO in the region 1. Around $kd = \pi/2$, ZT occurs, and part of light beam tunnels to the upper band and into the region 2, where it undergoes DO. When the light beam reach the boundary of the parabolic confinement at the right side, it can also tunnel to the lower band through ZT and undergoes RBO. In this way, the light beam undergoes the process LBO-ZT-DO-ZT-RBO. Figures \ref{fig:bxkPi}(e) and \ref{fig:bxkPi}(f) show the superposition of LBO, ZT, and RBO in the spatial space and reciprocal space, respectively. By comparing Fig.~\ref{fig:bxk0}(e) and Fig.~\ref{fig:bxkPi}(e), we can see that the spatial oscillation range of BDZO is much larger than that of BZO. When $\delta V = 1.0$, the band gap is too large to make the obvious occurrence of ZT. The light beam just undergo LBO, as shown in Figs.~\ref{fig:bxkPi}(h) and \ref{fig:bxkPi}(i).

It is known that the band structure can be reconstructed visually through a dispersion spectroscopic approach via the spatial evolution of light beam in the photonic lattices \cite{PhysRevLett.96.023901.2006}. Figure \ref{fig:FEA-HO}(a) shows the comparison between the rescaled spatial evolution of LBO and the band structure [as shown by solid lines in Fig.~\ref{fig:BPOWA}(b)]. By rescaling the propagation distance in unit of $Z_{BO}/(2\pi)$ and the waveguide index in unit of $N_h/a$ and shifting the waveguide index by $-1$ unit, the contour plots of spatial evolution of LBO match with the dispersion relation curves (solid lines).

The trajectories of BDZO can be roughly calculated through the Hamiltonian optics method. From the position-dependent
dispersion relation Eq.~(\ref{eq:Disp}), the evolution of the beam can be solved by using the equations of motion
\begin{equation}\label{eq:HO}
\frac{dx}{dz} = \frac{\partial \beta(x, k)}{\partial k}\,, \qquad
\frac{dk}{dz} = -\frac{\partial \beta(x, k)}{\partial x}\,.
\end{equation}
From Eq.~(\ref{eq:HO}), the period of BO is estimated as $Z_{BO} = 2\pi/|\partial \beta/\partial x|=\pi/|\alpha x|$. Since the space $z$ in OWA is analogous to the time $t$ in quantum system, the propagation period of light beam along the $z$ direction is the same as the oscillation period. The field-evolution analysis results [contour plots in Fig.~\ref{fig:FEA-HO}] can be fitted with the Hamiltonian optics results [solid line (LBO), dashed line (DO) and dotted line (RBO) in Fig.~\ref{fig:FEA-HO}]. We need to note that these Hamiltonian optics results are calculated by ignoring the multipole tunneling. In fact, due to the existence of the multipole tunneling, the pattern of the the field evolution is complicated. The trajectories from Hamiltonian optics method are used to confirm the superposition of LBO, DO and RBO. However, it is very hard to calculate ZT by using the Hamiltonian optics method, which can not show the variation of intensities. We may need other methods to obtain the tunneling rate of ZT in the Hamiltonian optics calculation.

\section{DISCUSSION AND CONCLUSION}

Thanks to the parabolic confinements in BPOWA, the superposition of BO, DO and ZT is realized. Regarding the experimental realization of such BPOWA, we can resort to the design of binary optical superlattices \cite{PhysRevLett.102.076802.2009} and parabolic optical waveguide arrays \cite{JOptSocAmB.27.1299.2010}. The two kinds of waveguides with different onsite propagation constants are placed alternatively. The parabolic confinement and offsets between them are realized by careful designing the structures, which is similar to the experimental design in AlGaAs waveguide arrays \cite{PhysRevLett.83.4756.1999}. The precise adjustment of each individual waveguide can be realized in nematic liquid crystals, which has a sensitive voltage-controlled flexibility \cite{Opto-Electon.Rev.15.210.2007}.

As mentioned \cite{PhysRevLett.102.076802.2009}, in the binary optical superlattices with linear confinement, both synchronous and asynchronous BZO are observed. However, in BPOWA, BO is approximate, and the period of DO depends on the position and propagation constant of input beam, as a result, it is almost impossible to realize synchronous BDZO. We can say that BDZO in BPOWA is asynchronous, except that we can make the period of BO $Z_{\rm BO}$ and that of DO $Z_{\rm DO}$ to fulfill the relation $Z_{\rm BO}/Z_{\rm DO} = p/q$, where $p$ and $q$ are integers.

In conclusion, we found that the propagation and tunneling of light beam in the binary parabolic optical waveguide arrays have close relation with its band structure. The band gap between the two minibands appears at the zone edge and can be tuned in width through the difference of the onsite propagation constants of two types of waveguides. Various combinations of optical oscillations and tunneling (BO, DO and ZT) are formed as demonstrated on the phase diagram. For certain band gap, both the Bloch-Zener oscillation (BZO) and the Bloch-dipole-Zener oscillation (BDZO) are realized with proper input light beams, which is demonstrated by the field-evolution analysis. Since the spatial oscillation range of BDZO is much larger than that of BZO, BDZO may have potential applications in optical splitting and waveguiding devices.

\section*{ACKNOWLEDGMENTS}

This work was supported by RGC General Research Fund of the
Hong Kong SAR Government.

\newpage

%\section*{REFERENCES}
% Bibliography
%\bibliographystyle{phaip}
%\pagestyle{plain}
%\bibliography{E:/Research/Biblibrary/biblib}

\clearpage

\section*{List of Figure Captions}

\noindent Fig. 1. (Color online) (a) Schematic diagram of a binary parabolic optical waveguide arrays (BPOWA). (b) Dispersion relation of unitary POWA (dotted lines) and binary POWA (solid lines). The parameters are $N=101$, $\alpha = 2.0$, $\kappa = 1.0$, and $\delta V = 0.5$. Fig1.eps

\noindent Fig. 2. (Color online) Phase diagram of BPOWA. There are two bands, whose boundaries are $\beta_{-}(x, 0)$, $\beta_{-}(x, \pi/2)$, $\beta_{+}(x, \pi/2)$, and $\beta_{+}(x, 0)$. Separated by three critical lines $\beta_{-}(0, \pi/2)$, $\beta_{+}(0, \pi/2)$, and $\beta_{+}(0,0)$, ten regions are formed. The correspondences between these regions and different optical oscillations are as follows. DO (dipole oscillation): 2 and 4. LBO (left Bloch oscillation): 1, 5, 7, and 9. RBO (right Bloch oscillation): 3, 6, 8, and 10. The parameters are $N=101$, $n_0=20$, $k_0 = \pi$, $\sigma=1$, $\alpha = 2.0$, $\kappa = 1.0$, and $\delta V = 0.5$. Fig2.eps

\noindent Fig. 3. (Color online) Band structures with components of input Gaussian beams, contour plots of $|\psi(x,z)|^2$ and $|\phi(k,z)|^2$ for three cases: (a)-(c) $\delta V = 0.0$, (d)-(f) $\delta V = 0.5$, and (g)-(i) $\delta V = 1.0$. The other parameters are $N=101$, $n_0=20$, $k_0 = 0$, $\sigma=1$, $\alpha = 2.0$, $\kappa = 1.0$. Fig3.eps

\begin{figure}[htbp]
  \centering
  \includegraphics[width=0.8 \textwidth]{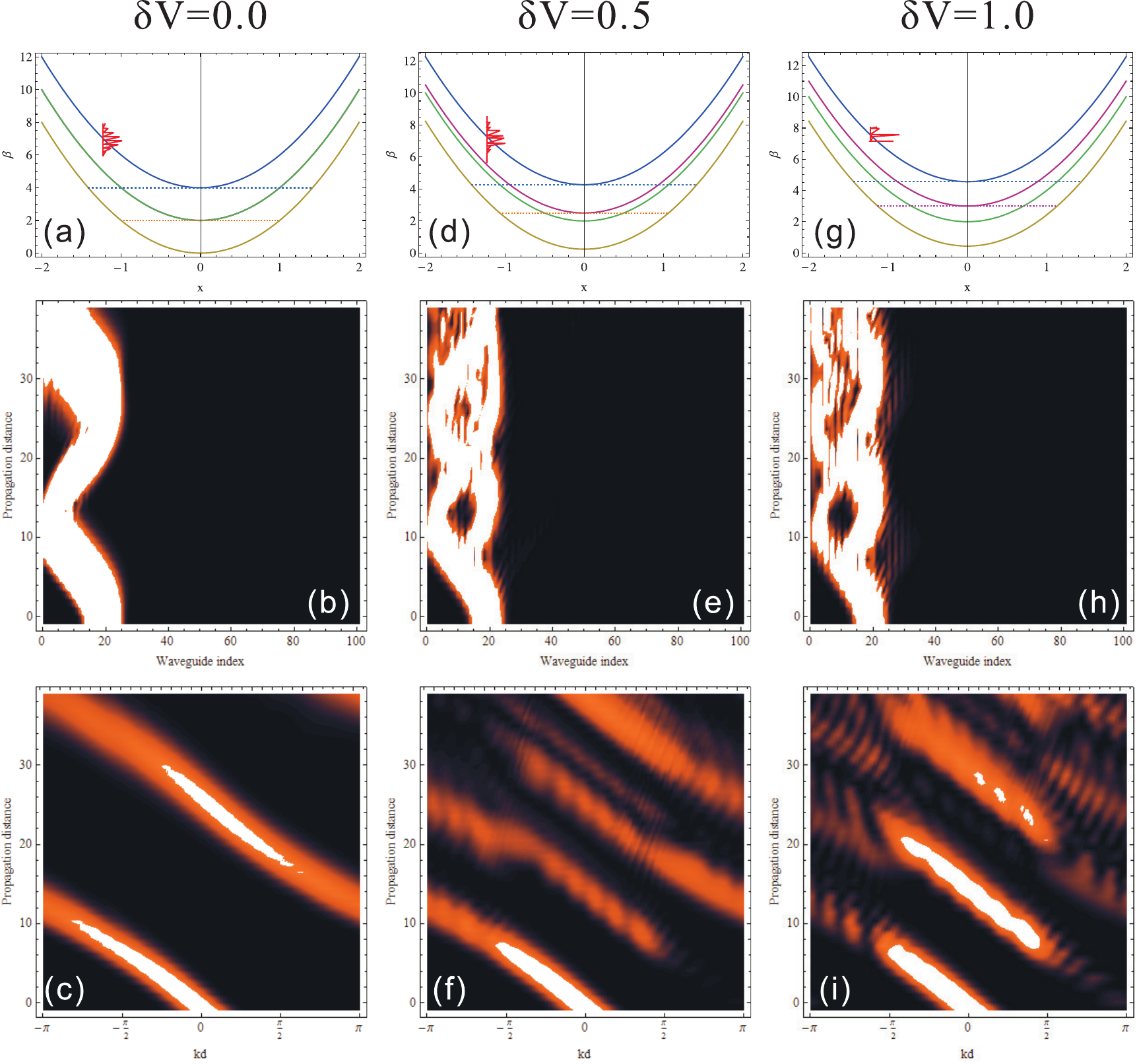}
  \caption{(Color online) Band structures with components of input Gaussian beams, contour plots of $|\psi(x,z)|^2$ and $|\phi(k,z)|^2$ for three cases: (a)-(c) $\delta V = 0.0$, (d)-(f) $\delta V = 0.5$, and (g)-(i) $\delta V = 1.0$. The other parameters are $N=101$, $n_0=20$, $k_0 = 0$, $\sigma=1$, $\alpha = 2.0$, $\kappa = 1.0$. Fig3.eps.}
  \label{fig:bxk0}
  \end{figure}

\noindent Fig. 4. (Color online) Band structures with components of input Gaussian beams, contour plots of $|\psi(x,z)|^2$ and $|\phi(k,z)|^2$ for three cases: (a)-(c) $\delta V = 0.0$, (d)-(f) $\delta V = 0.5$, and (g)-(i) $\delta V = 1.0$. The other parameters are the same as those in Fig.~\ref{fig:bxk0} except $k_0 = \pi$. Fig4.eps.

\begin{figure}[htbp]
  \centering
  \includegraphics[width=0.8 \textwidth]{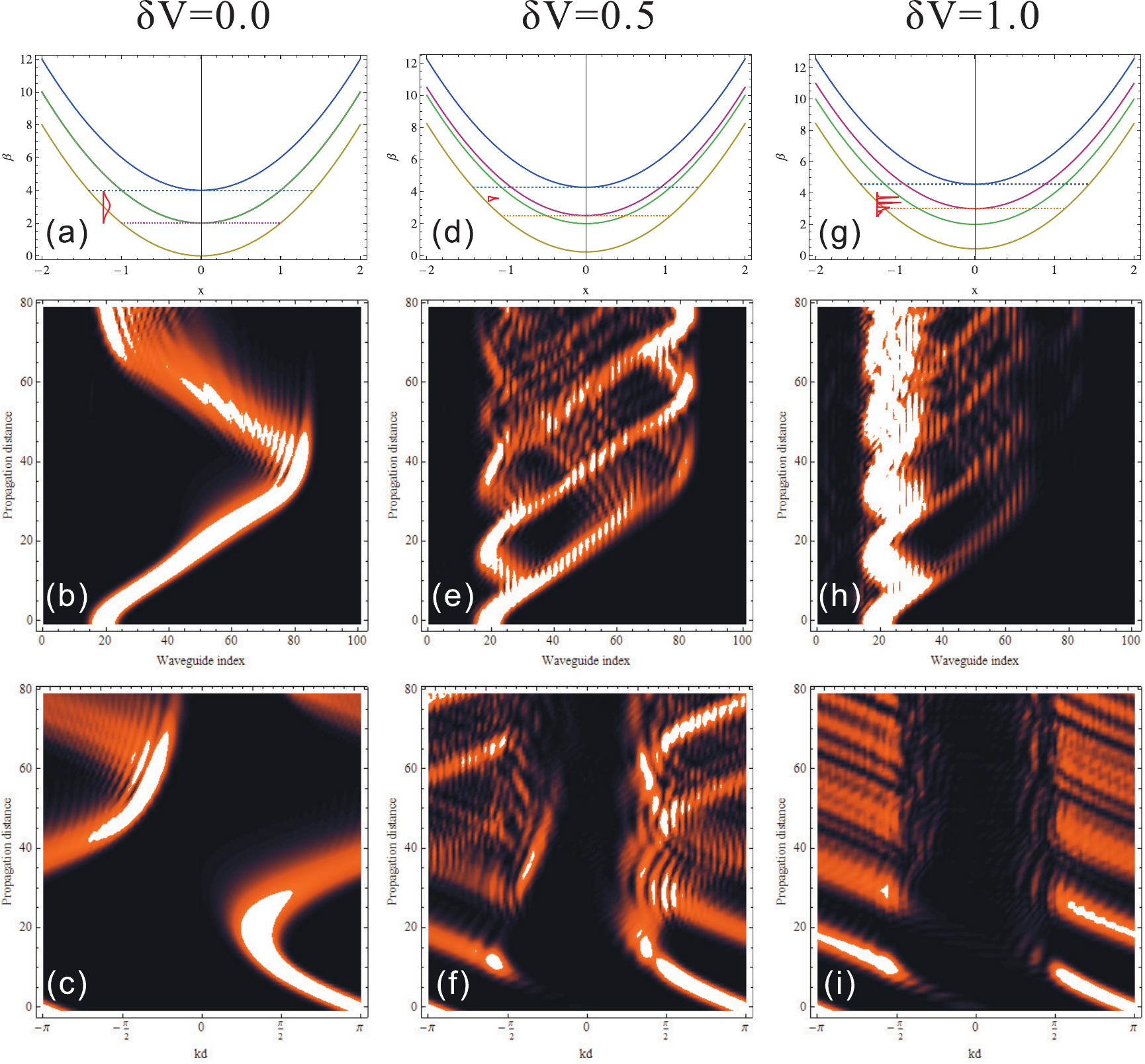}
  \caption{(Color online) Band structures with components of input Gaussian beams, contour plots of $|\psi(x,z)|^2$ and $|\phi(k,z)|^2$ for three cases: (a)-(c) $\delta V = 0.0$, (d)-(f) $\delta V = 0.5$, and (g)-(i) $\delta V = 1.0$. The other parameters are the same as those in Fig.~\ref{fig:bxk0} except $k_0 = \pi$. Fig4.eps.}
  \label{fig:bxkPi}
  \end{figure}

\begin{figure}[htbp]
  \centering
  \includegraphics[width=0.5 \textwidth]{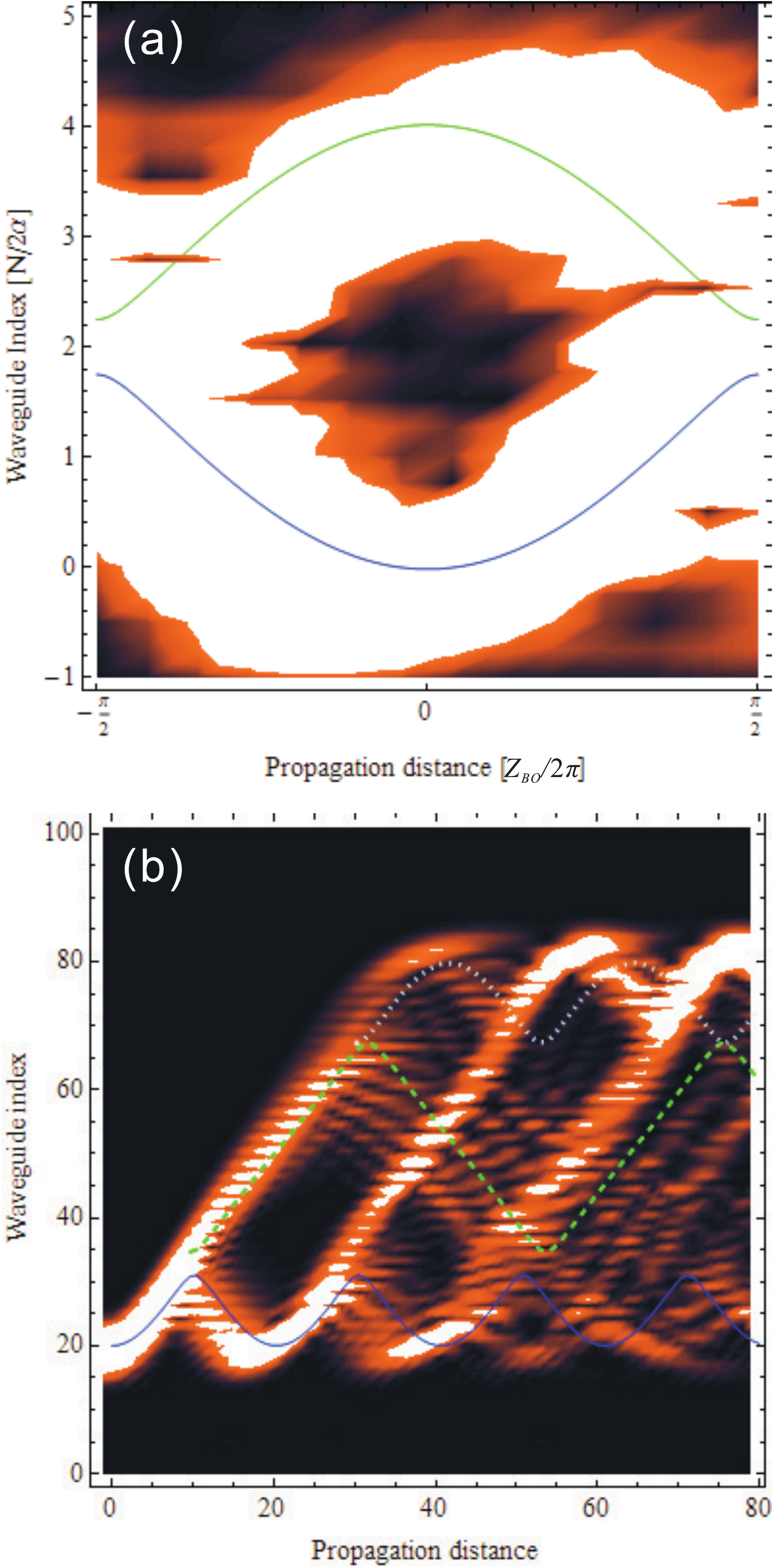}
  \caption{(Color online) (a) Visual demonstration of band structure through spatial evolution of light beam (left Bloch oscillation) in BPOWA. The parameters are the same as those in Fig.~\ref{fig:bxk0}(e). (b) Comparison of the field-evolution analysis result and Hamiltonian optics result of BDZO. The parameters are the same as those in Fig.~\ref{fig:bxkPi}(e). Fig5.eps.}
  \label{fig:FEA-HO}
  \end{figure}

\noindent Fig. 5. (Color online) (a) Visual demonstration of band structure through spatial evolution of light beam (left Bloch oscillation) in BPOWA. The parameters are the same as those in Fig.~\ref{fig:bxk0}(e). (b) Comparison of the field-evolution analysis result and Hamiltonian optics result of BDZO. The parameters are the same as those in Fig.~\ref{fig:bxkPi}(e). Fig5.eps.

\newpage

\newpage

\newpage

\newpage

\newpage

\end{document}